\documentclass[preprintnumbers,amsmath,amssymb]{revtex4}
\usepackage{graphicx}
\usepackage{amsmath, amsthm, amssymb,bm}
\begin{document}
\title{Hamiltonian Derivations of the Generalized Jarzynski Equalities under Feedback Control}

\author{Takahiro Sagawa$^{1,2}$}

\affiliation{$^1$ The Hakubi Center, The Kyoto University, Yoshida-Ushinomiya-cho, Sakyo-ku, Kyoto 606-8302, Japan \\
$^2$ Yukawa Institute of Theoretical Physics, The Kyoto University, Kitashirakawa Oiwake-Cho, 606-8502 Kyoto, Japan}


\begin{abstract}
In the presence of feedback control by ``Maxwell's demon,'' the second law of thermodynamics and the nonequilibrium equalities such as the Jarzynski equality need to be generalized.   In this paper, we derive the generalized Jarzynski equalities for classical Hamiltonian dynamics based on the Liouville's theorem, which is the same approach as the original proof of the Jarzynski equality [Phys.~Rev.~Lett.~\textbf{78}, 2690 (1997)].  The obtained equalities lead to the generalizations of the second law of thermodynamics for the Hamiltonian systems in the presence of  feedback control.
\end{abstract}

\maketitle

\section{Introduction}

Due to the recent advancement of nonequilibrium statistical mechanics, it has been established that the second law of thermodynamics can be expressed not only in terms of inequalities but also in terms of equalities, such as the fluctuation theorem and the Jarzynski equality~\cite{Evans,Gallavotti,Evans2,Crooks,Jarzynski2,Jarzynski1,Kawai}.  These equalities are universally valid even if the state of the system is  far from equilibrium. For the case of classical dynamics, the Jarzynski equality~\cite{Jarzynski1} is expressed as
\begin{equation}
\langle e^{-\beta (W - \Delta F)} \rangle = 1,
\label{Jarzynski}
\end{equation}
where $\langle \dots \rangle$ describes the ensemble average for all microscopic trajectories, $T = (k_{\rm B}\beta)^{-1}$ is the temperature of the initial canonical distribution with $k_{\rm B}$ being the Boltzmann constant,  $W$ is the work performed on the system, and $\Delta F$ is the free-energy difference.  We note that Eq.~(\ref{Jarzynski}) holds if the initial distribution of the system is  the canonical distribution.  The usual expression of the second law of thermodynamics is a straightforward consequence of the Jarzynski equality: by using the concavity of the exponential function, we obtain
\begin{equation}
\Delta F - \langle W \rangle \leq  0.
\label{second}
\end{equation}
Inequality~(\ref{second}) implies that  the amount of the free-energy gain of the system cannot be larger than the amount of the  energy-input (work) on the system on average.

However, if a thermodynamic system is subject to feedback control (or equivalently, the system is controlled by ``Maxwell's demon''~\cite{Maxwell,Szilard,Demon}), the second law of thermodynamics~(\ref{second}) needs to be generalized~\cite{Lloyd,Touchette,Cao1,Lopez,Sagawa-Ueda1,Cao2,Cao3,Maruyama,Suzuki1,Brandes,SWKim}. Here, ``feedback'' means  that the control protocol depends on the outcomes obtained by measurements.  For example, for the case of the Szilard engine~\cite{Szilard}, the demon can extract $k_{\rm B}T \ln 2$ of work from an isothermal cycle by using $1$ bit ($= \ln 2$ nat) of information.  The essential of the role of the demon is to increase the system's free energy with feedback control by using the information about the system.  Recently,  the Szilard-type Maxwell's demon has been experimentally realized by using colloidal particles~\cite{Toyabe}.

The Jarzynski equality~(\ref{Jarzynski}) also needs to be generalized in the presence of feedback control.  
In a previous work~\cite{Sagawa-Ueda3}, we have derived the generalizations of the Jarzynski equality for  classical stochastic dynamics based on the detailed fluctuation theorem~\cite{Crooks,Jarzynski2}.  Moreover, several nonequilibrium equalities with feedback control have also been derived for different setups~\cite{Kim,Suzuki2,Ponmurugan,Horowitz,Morikuni}, and relevant issues have been studied~\cite{Abreu,Horowitz2}.  We note that one of the generalized Jarzynski equalities  has been experimentally verified~\cite{Toyabe}.   However,  the derivations of the generalized Jarzynski equalities based on the classical Hamiltonian dynamics have not been studied yet.   In this paper, we will derive two generalizations of the Jarzynski equality based on the Hamiltonian dynamics, in particular based on the Liouville's theorem.

\section{Setup}

We consider a $d$-dimensional Hamiltonian system with $N$ classical  particles.  Let $x := (\bm r, \bm p)$ be the phase-space point of the system, where $\bm r$ and $\bm p$ respectively denote the $dN$-dimensional position and momentum.  We write the time-reversal of $x$ as $x^\ast := (\bm r, -\bm p)$. The Hamiltonian of the system is given by $H(x, \lambda)$, where $\lambda$ denotes external control parameters such as the frequency of the optical tweezers. 

We consider the dynamics of the system from time $t=0$ to $\tau$.
We drive the system from the thermal equilibrium by changing $\lambda$.  
 Let $x_t := x(t)$ and $\lambda_t := \lambda (t)$. The initial distribution of the system is assumed to obey the canonical distribution corresponding to $H (x_0, \lambda_0)$, whose probability density is given by
\begin{equation}
P_{\rm can} [x_0, \lambda_0] := \frac{e^{- \beta H(x_0, \lambda_0)}}{Z_0},
\end{equation}
where $Z_0 := \int dx_0 e^{-\beta H(x_0 , \lambda_0)}$ is the partition function.  The corresponding free energy is  $F_0 := -k_{\rm B}T \ln Z_0$.
The equation of motion is given by
\begin{equation}
\begin{split}
\frac{d \bm r (t)}{dt} &= \frac{\partial H}{\partial \bm p} (x(t), \lambda (t)), \\
\frac{d \bm p (t)}{dt} &= - \frac{\partial H}{\partial \bm r} (x(t), \lambda (t)). 
\end{split}
\label{eq1}
\end{equation}
We write the formal solution of Eq.~(\ref{eq1}) as
\begin{equation}
x_t =  M_t [x_0],
\end{equation}
where $M_t$ is a bijective map acting on the phase space.
The probability density of $x_t$, denoted as $P[x_t]$, is then given by
\begin{equation}
P[x_t] = P_{\rm can} [ M_t^{-1} (x_t), \lambda_0 ],
\end{equation}
where $M_t^{-1}$ is the inverse of $M_t$, and we used the Liouville's theorem.

We perform a measurement on the system at time $t_{\rm m}$ ($0 \leq t_{\rm m} < \tau$) and obtain outcome $y$.  We assume that the measurement involves a stochastic error  characterized by 
\begin{equation}
P [y  | x_{\rm m}],
\label{error}
\end{equation}
which describes the probability density of obtaining $y$ under the condition that the true state of the system is given by $x_{\rm m} := x(t_{\rm m})$ at time $t_{\rm m}$.  If $P [y  | x_{\rm m}] = \delta (y - x_{\rm m})$ holds, the measurement is error-free.  On the other hand, if the noise is Gaussian, the conditional probability is given by $P[y | x_{\rm m}] \propto \exp \left( - (y - x_{\rm m})^2 / (2N)  \right)$, where $N$ is the intensity of the noise.  
In general, the joint probability density of $x_{\rm m}$ and $y$ is given by
\begin{equation}
P [x_{\rm m}, y] = P [y  | x_{\rm m}]P[x_{\rm m}],
\end{equation}
and the probability density of obtaining $y$ is given by
\begin{equation}
P[y] = \int dx_{\rm m} P [x_{\rm m}, y].
\end{equation}
We note that $y^\ast$ denotes  time-reversal of $y$ corresponding to the time-reversal $x_{\rm} \mapsto x_{\rm}^\ast$.

We then discuss the concept of the mutual information~\cite{Cover-Thomas}, which is the key quantity to characterize the information content that is obtained by the measurement.
The mutual information, denoted as $\langle I \rangle$, is defined as 
\begin{equation}
\langle I \rangle := \int dx_{\rm m} dy P[x_{\rm m}, y] \ln \frac{P[x_{\rm m}, y]}{P[x_{\rm m}]P[y]},
\label{mutual}
\end{equation}
where we write
\begin{equation}
I[x_{\rm m}, y] := \ln \frac{P[x_{\rm m}, y]}{P[x_{\rm m}]P[y]}.
\end{equation}
The mutual information characterizes the correlation between state $x_{\rm m}$  and outcome $y$;  the more information we get, the larger $\langle I \rangle$ is.  In other words, the larger the error is, the less $\langle  I \rangle$  is.  In fact, if two probability variables $x_{\rm m}$ and $y$ are independent (i.e., $P[x_{\rm m}, y] = P[x_{\rm m}]P[y]$ holds), the mutual information vanishes so that $I[x_{\rm m}, y]  = 0$  holds for all $x_{\rm m}$ and $y$.

After the measurement at time $t_{\rm m}$, we perform feedback control on the system so that the control protocol of  $\lambda$ depends on outcome $y$ at time $t$ ($> t_{\rm m}$).  To explicitly express the effect of the feedback, we write
\begin{equation}
x_t(y) = M_t(y) [x_0] \ (t > t_{\rm m}),
\label{eq2}
\end{equation}
where $x_t(y)$ is the state of the system at time $t$ ($> t_{\rm m}$), under the condition that the initial state  is $x_0$ and the measurement outcome is $y$.  
Equality~(\ref{eq2}) means that the Hamiltonian evolution of the system is determined by outcome $y$ after  time $t_{\rm m}$.
We note that  map $M_t(y)$ satisfies  the Liouville's theorem for each $y$.

The Hamiltonian at time $\tau$ may also depend on outcome $y$ as $H (x_\tau(y), \lambda_\tau(y))$, where $\lambda_\tau(y)$ denotes the value of external parameter $\lambda$ at time $\tau$ under the condition that the outcome is given by $y$.  The corresponding canonical distribution is given by 
\begin{equation}
P_{\rm can} [x_\tau(y), \lambda_\tau(y)] := \frac{e^{- \beta H(x_\tau(y), \lambda_\tau(y))}}{Z_\tau(y)},
\end{equation}
where $Z_\tau(y) := \int dx_\tau(y) e^{-\beta H(x_\tau(y) , \lambda_\tau(y))}$ is the partition function which gives  the free energy: 
\begin{equation}
F_\tau(y) := -k_{\rm B}T \ln Z_\tau(y).
\end{equation}
 We note that the probability distribution of $x_\tau(y)$ is not necessarily given by the canonical distribution; $P[x_\tau(y)] = P_{\rm can} [x_\tau(y), \lambda_\tau(y)]$ does not necessarily hold.

In the total process, the work performed on the system is given by
\begin{equation}
W :=  H(x_\tau(y), \lambda_\tau(y)) - H(x_0, \lambda_0) = H(M_\tau(y) [x_0], \lambda_\tau(y)) - H(x_0, \lambda_0),
\end{equation}
which is determined by initial state $x_0$ and outcome $y$. On the other hand, the free-energy difference is given by
\begin{equation}
\Delta F := F_\tau(y) - F_0,
\label{free_energy}
\end{equation}
which depends on outcome $y$.

\section{Main Results}

We now derive the two generalizations of the Jarzynski equality.  The first one is given by  
\begin{equation}
\langle e^{-\beta (W - \Delta F) - I} \rangle = 1,
\label{main1}
\end{equation}
where $\langle \cdots \rangle$ describes the ensemble average with respect to $x_0$ and $y$.  In fact, $W$ is determined by $x_0$ and $y$, $\Delta F$ by $y$, and $I$ by $y$ and $x_0$ through $x_{\rm m} = M_{t_{\rm m}} [x_0]$.
The key feature of Eq.~(\ref{main1}) is that the left-hand side involves the term of the mutual information obtained by the measurement.

The proof of Eq.~(\ref{main1}) is as follows.  We first assume that $P [y  | x_{\rm m}] \neq 0$ holds for all $y$ and $x_{\rm m}$. By using the joint distribution $P[x_0, y]$, we obtain
\begin{equation}
\begin{split}
\langle e^{-\beta (W - \Delta F)- I}  \rangle &= \int dx_0 dy \frac{e^{-\beta H(x_0, \lambda_0)}}{Z_0}  P[y | x_{\rm m}] e^{-\beta (H(x_\tau(y), \lambda_\tau(y)) - H(x_0, \lambda_0)) } \frac{Z_0}{Z_\tau(y)} \frac{P[y]}{P[y | x_{\rm m}]}  \\
&=   \int dx_\tau(y) dy \frac{e^{-\beta H(x_\tau(y), \lambda_\tau(y))}}{Z_\tau(y)}  P[y], 
\end{split}
\end{equation}
where we used the Liouville's theorem $dx_0 = dx_\tau(y)$.  By noting that $\int dx_\tau(y)  e^{-\beta H(x_\tau(y), \lambda_\tau(y))} = Z_\tau(y)$, we obtain Eq.~(\ref{main1}).  We note that Eq.~(\ref{main1}) has been obtained in Ref.~\cite{Sagawa-Ueda3} for classical stochastic systems.

By using the concavity of the exponential function, we have
\begin{equation}
\langle e^{-\beta (W - \Delta F) - I} \rangle \geq e^{- \langle \beta (W - \Delta F) - I \rangle}.
\end{equation}
Therefore, Eq.~(\ref{main1}) leads to 
\begin{equation}
\langle  \Delta F - W \rangle \leq k_{\rm B} T \langle I \rangle, 
\label{main_inequality1}
\end{equation}
which is the generalized second law of thermodynamics.
Inequality  (\ref{main_inequality1}) implies that, by using feedback control,  the free-energy increase can be lager than the performed work by the term proportional to the mutual information obtained by the measurement.
The equality in (\ref{main_inequality1}) is achieved by the Szilard engine~\cite{Szilard}, where $\langle \Delta F \rangle = 0$, $\langle W \rangle = - k_{\rm B}T \ln 2$, and $\langle I \rangle = \ln 2$ hold.
We note that inequality (\ref{main_inequality1}) has been obtained in Ref.~\cite{Sagawa-Ueda1} for quantum systems and in Ref.~\cite{Sagawa-Ueda3} for classical stochastic systems.

We next discuss the second generalization of the Jarzynski equality, which is given by
\begin{equation}
\langle e^{-\beta (W - \Delta F ) } \rangle = \gamma,
\label{main2}
\end{equation}
where $\gamma$ characterizes the efficacy of feedback control.  In fact, $\gamma$ is quantitatively defined as follows.  We consider the backward or time-reversed process of the feedback control with outcome $y$.  The initial state of the backward process is given by the canonical distribution corresponding to $\lambda_\tau (y)$, and  the backward control protocol is given by $\lambda_t^\dagger (y) := \lambda_{\tau - t} (y)$ for each $y$ that is obtained in the forward process.  We note that we do not perform any feedback in the backward process. We then perform a measurement on the system at time $\tau - t_{\rm m}$ during the backward process, and obtain outcome $y'$. We note that $y'$ does not necessarily  equal to $y^\ast$.  We write as $P^\dagger_{(y)} [y']$ the probability density of obtaining $y'$ by the measurement during the backward process with control protocol $\lambda_t^\dagger (y)$.  In particular, the probability density of $y' = y^\ast$  is written as $P^\dagger_{(y)} [y^\ast]$.  Then, $\gamma$ is defined as
\begin{equation}
\gamma := \int dy P^\dagger_{(y)} [y^\ast].
\label{gamma}
\end{equation}
Parameter $\gamma$ characterizes how efficiently the feedback is performed, in terms of the sum of the probabilities that the time-reversed outcome $y^\ast$ is obtained during the backward process with time-reversed protocol $\lambda_t^\dagger (y)$.  We note that  $\gamma = 1$ holds without feedback control, because $\{ P^\dagger [y^\ast] \}$ becomes a single probability distribution for such cases.  On the other hand, $\gamma = 2$ holds for the case  the Szilard engine, where the number of the outcomes is two and the feedback control is perfect~\cite{Sagawa-Ueda3}.  We note that Eq.~(\ref{main2}) has been experimentally verified for a classical stochastic system~\cite{Toyabe}.
 
We now prove Eq.~(\ref{main2}).  We first obtain 
\begin{equation}
\begin{split}
\langle e^{-\beta (W - \Delta F)}  \rangle &= \int dx_0 dy \frac{e^{-\beta H(x_0, \lambda_0)}}{Z_0}  P[y | x_{\rm m}] e^{-\beta (H(x_\tau(y), \lambda_\tau (y)) - H(x_0, \lambda_0)) } \frac{Z_0}{Z_\tau(y)}  \\
&=   \int dx_0 dy \frac{e^{-\beta H(x_\tau(y), \lambda_\tau(y))}}{Z_\tau(y)}  P[y | x_{\rm m}] \\
&= \int dx_0 dy \frac{e^{-\beta H(x_\tau^\ast (y), \lambda_\tau(y))}}{Z_\tau(y)}  P[y^\ast | x_{\rm m}^\ast ],
\end{split}
\end{equation}
where we assumed that the error is time-reversal symmetric as
\begin{equation}
P [y | x_{\rm m}] = P [y^\ast | x_{\rm m}^\ast],
\end{equation}
and that the Hamiltonian is also time-reversed symmetric as $H(x_\tau(y), \lambda_\tau (y)) =H(x_\tau^\ast (y), \lambda_\tau (y))$.
By noting that the Liouville's theorem $dx_0 = dx_\tau(y)$ holds for each $y$, and  noting that $dx_\tau(y) = dx_\tau^\ast (y)$ holds, we have
\begin{equation}
\langle e^{-\beta (W - \Delta F)}  \rangle = \int dx_\tau^\ast (y) dy \frac{e^{-\beta H(x_\tau^\ast (y))}}{Z_\tau(y)}  P[y^\ast | x_{\rm m}^\ast ].
\end{equation}
Since the Hamiltonian dynamics is reversible,  $x_{\rm m}^\ast = [M_{\tau - t_{\rm m}}(y)]^{-1}[x_\tau^\ast (y)]$ holds.  Therefore, we obtain
\begin{equation}
\langle e^{-\beta (W - \Delta F)}  \rangle =  \int dx_\tau^\ast (y) dy \frac{e^{-\beta H(x_\tau^\ast (y))}}{Z_\tau(y)}  P[y^\ast | [M_{\tau - t_{\rm m}}(y)]^{-1}(x_\tau^\ast (y))] = \int dy P^{\dagger}_{(y)} [y^\ast] = \gamma,
\end{equation}
which proves Eq.~(\ref{main2}).   
We note that Eq.~(\ref{main2}) has been derived for classical stochastic systems~\cite{Sagawa-Ueda3} and for quantum systems~\cite{Morikuni}.
We also note that Eq.~(\ref{main2}) is a straightforward consequence of a result in Ref.~\cite{Kawai} for  a special class of  measurements.
Equality~(\ref{main2}) leads to
\begin{equation}
\langle  \Delta F - W \rangle \leq k_{\rm B} T \ln \gamma. 
\label{main_inequality2}
\end{equation}
The equality in (\ref{main_inequality2}) is achieved for the case of the Szilard engine where $\gamma = 2$ holds.

We now discuss the relationship between the two generalizations of the Jarzynski equality.  The first one (\ref{main1}) only involves the term of the information $I$ obtained by the measurement, which is independent of the protocol of the feedback control.  On the other hand, the second one (\ref{main2}) involves the term $\gamma$ that characterizes the efficacy of feedback control, which describes how efficiently the information is used by the feedback.  We then discuss the quantitative relationship between $I$ and $\gamma$.   We introduce notation $C[X] := \ln \langle e^{-X} \rangle$ for arbitrary probability variable $X$.  We note that $C[I] = 0$ holds.  Then, from Eqs.~(\ref{main1}) and (\ref{main2}), we obtain 
\begin{equation}
C[\sigma + I ] - C[\sigma] - C[I] = - \ln \gamma,
\label{rel1}
\end{equation}
where $\sigma := \beta (W - \Delta F)$.  If the joint probability distribution of $\sigma$ and $I$ is Gaussian, Eq.~(\ref{rel1}) reduces to
\begin{equation}
\langle \sigma I \rangle - \langle \sigma \rangle \langle I \rangle = - \ln \gamma.
\label{rel2}
\end{equation}
Eqs.~(\ref{rel1}) and (\ref{rel2}) imply that $\gamma$ characterizes the correlation between $\sigma$ and $I$; the larger  $\gamma$ is, the larger the efficiency of decreasing $\sigma$ by using $I$ is.

In the conventional thermodynamics without feedback control, the free energy is not a probability variable. On the other hand, with feedback control, the final free energy $F(y)$ can be a probability variable, because measurement outcome $y$  is a probability variable.  The free-energy difference $\Delta F$ defined in (\ref{free_energy}) then needs to be inside the statistical average $\langle \cdots \rangle$ in the generalizations of the second law of thermodynamics (\ref{main_inequality1}) and (\ref{main_inequality2}).  Therefore, the generalized second laws work only when we observe the ensemble of thermodynamic systems and take the ensemble average both in terms of phase-space point $x$ and outcome $y$.
 This is a characteristic of thermodynamics of feedback control.  If  the control protocol is independent of outcome $y$, the free-energy difference also becomes independent of $y$ and  inequalities (\ref{main_inequality1}) and (\ref{main_inequality2}) reduce to the conventional second law of thermodynamics.

In conclusion, we have derived the two generalizations of the Jarzynski equality, Eqs.~(\ref{main1}) and (\ref{main2}), based on the Hamiltonian dynamics.  The former involves the term of the obtained mutual information, and the latter involves the term of the feedback efficacy.  
 The key of the present derivations is the initial canonical distribution  and the Liouville's theorem.
The equalities lead to the two generalizations of the second law of thermodynamics (\ref{main_inequality1}) and (\ref{main_inequality2}), which give the fundamental bounds of the free-energy gain of thermodynamic systems that are subject to feedback control.  We note that our results are consistent with the second law of thermodynamics, if we take into account the energy cost needed for the controller during the measurement and the information erasure~\cite{Sagawa-Ueda2}.

\begin{acknowledgments}
TS acknowledges Prof. Masahito Ueda for a lot of valuable discussions.
This work was supported by Grants-in Aid for Scientific Research (KAKENHI 22103005 and 22340114), the
Global COE Program ``the Physical Sciences Frontier", and the Photon Frontier Network Program of MEXT of
Japan. TS also acknowledges support from JSPS (Grant  No. 208038).
\end{acknowledgments}

\end{document}